\documentclass[prl,aps,twocolumn,superscriptaddress,amssymb,floatfix,showpacs,amsfonts]{revtex4} 
\usepackage{graphicx,psfrag,amsmath,amssymb,float}
\usepackage{epstopdf}
\usepackage{color}
\input{epsf}
\usepackage{graphicx,natbib}

\newcommand{\bc}{\begin{center}}
\newcommand{\ec}{\end{center}}
\newcommand{\be}{\begin{equation}}
\newcommand{\ee}{\end{equation}}
\newcommand{\bea}{\begin{eqnarray}}
\newcommand{\eea}{\end{eqnarray}}

\def\12{\frac{1}{2}}

\makeatletter
\usepackage{psfrag}

\begin{document}

\title{Kondo Screening Cloud and Charge Quantization in Mesoscopic Devices}

\author{Rodrigo G. Pereira}

\affiliation{Department of Physics and Astronomy, University of British Columbia,
Vancouver, BC, Canada, V6T 1Z1}

\author{Nicolas Laflorencie}

\affiliation{Department of Physics and Astronomy, University of British Columbia,
Vancouver, BC, Canada, V6T 1Z1}
\affiliation{Insitute of Theoretical Physics, \'Ecole Polytechnique F\'ed\'erale de Lausanne, Switzerland}

\author{Ian Affleck}
\affiliation{Department of Physics and Astronomy, University of British Columbia,
Vancouver, BC, Canada, V6T 1Z1}

\author{Bertrand~I.~Halperin}
\affiliation{Physics Department, Harvard University, Cambridge, MA
02138, USA} 

\date{\today}

\begin{abstract}
We propose that the finite size of the Kondo screening cloud, $\xi_{K}$,
can be probed by measuring the charge quantization in a one-dimensional
system coupled to a small quantum dot. When the chemical potential
in the system is varied at zero temperature, one should observe charge
steps that are controlled by the Kondo effect when the system size
$L$ is comparable to $\xi_{K}$.  We show that, 
if the standard Kondo model is used, the ratio between the widths
of the Coulomb blockade valleys with odd or even number of electrons 
is a universal scaling function of $\xi_{K}/L$. 
\end{abstract}


\maketitle
The Kondo effect can be described as the strong renormalization of
the exchange coupling between an electron gas and a localized spin
at low energies \cite{hewson}. Below some characteristic energy scale,
known as the Kondo temperature $T_{K}$, a non-trivial many-body state
arises in which the localized spin forms a singlet with one conduction
electron. In the recent realizations of the Kondo effect~\cite{glazman}, 
a semiconductor
quantum dot in the Coulomb blockade regime plays the role of a spin
$S=1/2$ impurity when the number of electrons in the dot is odd.
The Kondo temperature in these systems is a function of the coupling
between the dot and the leads and can be conveniently controlled by
gate voltages. As a clear signature of the Kondo effect, one observes
that the conductance through a quantum dot with $S=1/2$ is a universal
scaling function $G\left(T/T_{K}\right)$ and reaches the unitary
limit $G=2e^{2}/h$ when $T\ll T_{K}$ for symmetric coupling to the leads \cite{delft}. On the other
hand, we expect that in a finite system the infrared singularities
that give rise to the Kondo effect are cut off not by temperature,
but by the level spacing $\Delta$ if $\Delta\gg T$. For a one-dimensional
(1D) system with length $L$ and characteristic velocity $v$, the relevant
dimensionless parameter is $\Delta/T_{K}\sim\xi_{K}/L$, where $\xi_{K}\equiv v/T_K$
is identified with the size of the Kondo screening cloud, the mesoscopic
sized wave function of the electron that surrounds and screens the
localized spin. This large length scale ($\xi_{K}\sim0.1-1\mu\textrm{m}$)
is comparable with the size of currently studied mesoscopic devices,
which has motivated several proposals that the Kondo cloud should
manifest itself through finite size effects in such systems \cite{kondobox,persistentcurrent,key-4}. 

One familiar property of small metallic islands in the Coulomb blockade
regime is the quantization of charge. Since each electron added to
the system costs a finite energy, the number of electrons changes
by steps as one varies the chemical potential and sharp conductance
peaks are observed at the charge degeneracy points \cite{glazman}.
 In AlGaAs-GaAs heterostructures, the level
spacing of a 1D system with size $L\sim1\mu$m is of
order $\Delta\sim100\mu eV$, large enough to be resolved experimentally.
Clearly, the energy levels should be affected by the Kondo interaction
with an adjacent dot. Therefore, the addition spectrum should exhibit
signatures of the finite size of the Kondo cloud. 

\psfrag{Vg}{$V_g$}

\psfrag{Vdw}{$V_{dw}$}

\psfrag{Phi}{$\Phi$}

\begin{figure}
\includegraphics[scale=0.5]{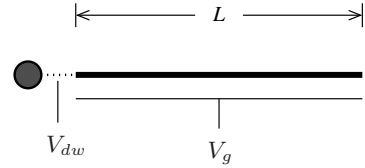}

\caption{Possible experimental setup. $V_{dw}$ controls the tunneling $t^{\prime}$ between
the small dot (on the left) and the wire and $V_{g}$ varies the chemical
potential in the wire. \label{cap:Possible-experimental-setup.}}
\end{figure}

We consider the geometry shown in Fig. \ref{cap:Possible-experimental-setup.}.
One of the ends of a wire of length $L$ with a large number of electrons
(such that $\Delta\ll\epsilon_{F}$, where $\epsilon_{F}$ is the
Fermi energy) is weakly coupled to a small quantum dot via a gate
voltage $V_{dw}$. Both the dot and the wire (which can be thought
of as a very long and thin dot) are in the Coulomb blockade regime.
The dot has a very large charging energy and a ground state
with $S=1/2$. We assume that the wire is very weakly connected to
a reservoir and that transfer of a single electron between wire
and reservoir could be measured by some Coulomb blockade technique.
This coupling is assumed weak enough so as not to affect the energy
levels of the wire-dot system. Experiments with small dots and large
dots have been performed, for example, in \cite{goldhaber-gordon}
and considered theoretically in \cite{key-4}.  

Here we investigate this effect using the basic Kondo model, ignoring 
charge fluctuations on the dot, assuming the wire contains 
a single channel and ignoring all electron-electron interactions 
in the wire. Within the constant interaction approximation \cite{glazman}
these interactions are represented by a simple charging energy whose 
effects on the charge steps are easily subtracted off. We postpone 
a more complete treatment of these complications, 
 to a later work. (The effect of short range interactions on 
the conductance through a Kondo impurity was treated in the Luttinger 
liquid framework in [\onlinecite{persistentcurrent}].)

\emph{Weak coupling}. -- The tunneling between the wire and the dot is well described by the Anderson model \cite{persistentcurrent}. For zero tunneling amplitude  the system reduces to an open chain and a free spin. In the continuum limit, we introduce the fermionic field $\Psi(x)$ for the electrons in the wire, as well as its right- and left-moving components \[
\Psi\left(x\right)=e^{ik_{F}x}\psi_{R}\left(x\right)+e^{-ik_{F}x}\psi_{L}\left(x\right).\] 
The open boundary conditions $\Psi\left(0\right)=\Psi\left(L\right)=0$
imply that $\psi_{L,R}$ are not independent. Instead, $\psi_{L}$
can be regarded as the extension of $\psi_{R}$ on the negative-$x$
axis: $\psi_{R}\left(-x\right)=-\psi_{L}\left(x\right)$. We can then
work with right movers only and write down an effective Kondo model (we drop the index $R$ hereafter)\begin{equation}
H=-iv_{F}\int_{-L}^{L}dx\,\psi^{\dagger}\partial_{x}\psi+2\pi v_{F}\lambda\psi^{\dagger}\left(0\right)\frac{\vec{\sigma}}{2}\psi\left(0\right)\cdot\vec{S},
\label{Hamil}
\end{equation}
where $\lambda$ is the dimensionless Kondo
coupling.  We assume that $\lambda$ is independent
of $V_g$, which may require that $t^{\prime}$ and the energy of the localized state in the dot
be tuned accordingly. For $\lambda\ll1$, the size of the Kondo screening
cloud (defined in the thermodynamic limit) is exponentially large: 
$\xi_{K}\sim(v_{F}/D)\, e^{1/\lambda}$, where $D\ll\epsilon_{F}$
is a high-energy cutoff. The boundary condition in the weak coupling
fixed point ($L/\xi_{K}\rightarrow0$) reads $\psi\left(-L\right)=e^{i\delta}\psi\left(L\right)$,
where $\delta=2k_{F}L$ mod $\pi$. Using the mode expansion $\psi\left(x\right)=-(i/\sqrt{2L})\sum_{k}e^{ikx}c_{k}$,
we obtain the momentum eigenvalues $k_{n}=\left(n\pi-\delta\right)/L,\, n\in\mathbb{Z}$. 

\begin{figure}
\includegraphics[%
  scale=0.33]{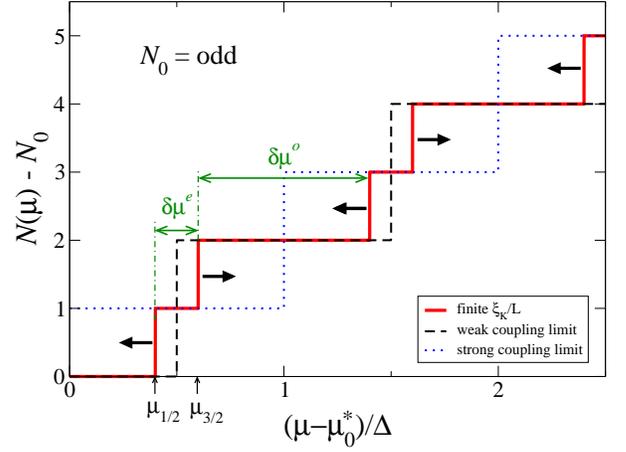}

\caption{(Color online) Charge quantization steps for the wire coupled to
a small quantum dot. The arrows indicate the direction in which the
single steps move as $\lambda_{eff}\left(L\right)$ grows.\label{cap:Charge-steps.}}
\end{figure}

We denote by $N(\mu )$ the total number of electrons in the system, including 
the one in the quantum dot.
 We are interested
in the elementary steps of $N$ around some initial value
$N_{0}\equiv N\left(\mu_{0}^*\right)$. At $T=0$, we calculate $N\left(\mu\right)$ as the integer value of $N$ that minimizes the thermodynamic potential
$\Omega\left(N\right)=E\left(N\right)-\mu N$, where $E\left(N\right)$
is the ground state energy. Defining $\mu_{\ell+\frac{1}{2}}$ as the value of $\mu$ where $N$ changes from $N_0+\ell$ to $N_0+\ell+1$, it follows  from the charge degeneracy condition $\Omega\left(N_{0}+\ell+1\right)=\Omega\left(N_{0}+\ell\right)$ that $\mu_{\ell+\frac{1}{2}}=E\left(N_{0}+\ell+1\right)-E\left(N_{0}+\ell\right)$.
For $\lambda=0$, we set $\mu_{0}^*$ halfway between the highest occupied
and the lowest unoccupied energy level of the open chain. In this
case, $N_{0}=$ odd: the ground state is doubly degenerate (total
spin $S_{tot}=1/2$) and consists of one electron in the dot and pairs
of conduction electrons in the Fermi sea. The energy levels measured
from $\mu_{0}^*$ are $\epsilon_{n}=\left(n-1/2\right)\Delta$, where
$\Delta=\pi v_{F}/L$. It is easy to see
that, as we raise $\mu>\mu_{0}$, $N$ jumps whenever $\mu$ crosses
an energy level. Moreover, there are only double steps due to the
spin degeneracy of the single-particle states ($\mu_{2m+\frac{1}{2}}=\mu_{2m+\frac{3}{2}}$).
As a result, $N\left(\mu\right)-N_{0}$ takes on only even
values and $N$ is always odd (see dashed line in Fig. \ref{cap:Charge-steps.}).

We obtain the charge steps in the weak coupling limit by calculating
the correction to the ground state energy using perturbation theory
in $\lambda$ \cite{persistentcurrent}. In the limit $\lambda\rightarrow0$,
the ground state for $N$ odd is still doubly degenerate and we write
$\left|GS\left(N=\textrm{odd}\right)\right\rangle =\prod_{n\leq0}c_{k_{n}}^{\uparrow\dagger}c_{k_{n}}^{\downarrow\dagger}\left|0\right\rangle \otimes\left|\gamma\right\rangle $,
where $\left|\gamma\right\rangle =\left|\Uparrow\right\rangle ,\left|\Downarrow\right\rangle $
is the spin state of the dot. However, for $N$ even, there
is one single electron occupying the state at the Fermi surface (with
momentum $k_{F}$). To lowest order in degenerate perturbation theory,
the ground state for $\lambda\rightarrow0$ is $\left|GS\left(N=\textrm{even}\right)\right\rangle =\prod_{n\leq0}c_{k_{n}}^{\uparrow\dagger}c_{k_{n}}^{\downarrow\dagger}\left|0\right\rangle \otimes\left|s\right\rangle $,
where $\left|s\right\rangle \equiv[\left|k_{F}\uparrow\right\rangle \otimes\left|\Downarrow\right\rangle -\left|k_{F}\downarrow\right\rangle \otimes\left|\Uparrow\right\rangle ]/\sqrt{2}$
is the singlet state between the spin of the dot and the electron
at $k_{F}$. This state has $S_{tot}=0$ and $\left\langle \right.\vec{S}\cdot\vec{S}_{el}\left.\right\rangle =-3/4$,
where $\vec{S}_{el}\equiv\sum_{k}c_{k}^{\dagger}\frac{\vec{\sigma}}{2}c_{k}^{\phantom{\dagger}}$
is the total spin of the conduction electrons. The singlet formation
lowers $E\left(N\right)$ for $N=$ even relatively to $N=$ odd.
As a result, the Kondo interaction splits the double steps and gives
rise to small plateaus with $N$ even (Fig. \ref{cap:Charge-steps.}).
To $O\left(\lambda^{2}\right)$, we find\begin{equation}
\frac{\mu_{\ell+\frac{1}{2}}-\mu_0^*}{\Delta}=\frac{\left[\ell\right]}{2}-\frac{3}{4}(-1)^\ell\left(\lambda+\lambda^{2}\ln\frac{DL}{v_{F}}+\dots\right),\label{eq:stepsweakcoup}\end{equation}
where $[\ell]\equiv\ell+1$ for $\ell$ even and $[\ell]\equiv\ell$ for $\ell$ odd. Note
the logarithmic divergence at $O\left(\lambda^{2}\right)$ as $L\rightarrow\infty$,
characteristic of the Kondo effect. We recognize the expansion of
the effective coupling $\lambda_{eff}\left(L\right)\sim\left[\ln\left(\xi_{K}/L\right)\right]^{-1}$
in powers of the bare $\lambda$, as expected from scaling arguments.
We have verified that the scaling holds up to $O\left(\lambda^{3}\right)$,
\emph{i.e}., $\mu_{\ell+\frac{1}{2}}$ has no dependence on the cutoff $D$ but
the implicit one in the expansion of $\lambda_{eff}\left(L\right)$.
This is remarkable given that $E\left(N\right)$ itself is cutoff
dependent. Based on this, we conjecture that the relative width
of the plateaus with $N$ odd/even is a universal scaling function
of $\xi_{K}/L$. We define the ratio (see Fig. \ref{cap:Charge-steps.})\begin{equation}
R\equiv\frac{\delta\mu^{\rm{o}}}{\delta\mu^{\rm{e}}}=\frac{E(N_0+3)-2 E(N_0+2)+E(N_0+1)}{E(N_0+2)-2 E(N_0+1)+E(N_0)}.\end{equation}
From Eq. (\ref{eq:stepsweakcoup}), we have that in the weak coupling
limit\begin{equation}
R\left(\frac{\xi_{K}}{L}\gg1\right)\approx\frac{2}{3}\ln\left(\frac{\xi_{K}}{L}\right)-\frac{1}{3}\ln \ln\left(\frac{\xi_K}{L}\right)+\textrm{const.}.
\label{WC}
\end{equation}

\emph{Strong coupling.} -- When $\lambda_{eff}\rightarrow\infty$,
the spin of the dot forms a singlet with one conduction electron and
decouples from the other electrons in the chain. In this limit the
Kondo cloud is small ($\xi_{K}/L\rightarrow0$). The strong coupling
boundary conditions reflect the $\pi/2$ phase shift for the particle-hole
symmetric case: $\psi\left(-L\right)=-\psi\left(L\right)$ \cite{comment}.
This implies that the $k$ eigenvalues are shifted with respect to
weak coupling: $k_{n}=(n\pi+\pi/2-\delta)/L$. At this fixed point
we recover the spin degeneracy of the single-particle states and double
steps in $N\left(\mu\right)$ appear at the shifted energy levels
($\epsilon_{n}=n\Delta$ as measured from the original $\mu_{0}^*$).
The ground state consists now of a singlet
plus pairs of electrons in the wire, thus $N\left(\mu\right)$ is
always even. Fig. \ref{cap:Charge-steps.} illustrates how the charge
staircase evolves monotonically from weak to strong coupling.

We explore the limit $k_{F}^{-1}\ll\xi_{K}\ll L$ by working out a
local Fermi liquid theory \cite{nozieres}. The idea is that virtual
transitions of the singlet at $x=0$ induce a local interaction in
the spin sector of the conduction electrons. The leading irrelevant
operator that perturbs the strong coupling fixed point and respects
SU(2) symmetry is \begin{equation}
H_{FL}=-\frac{2\pi^{2}}{3}\frac{v_{F}^{2}}{T_{K}}\left[\psi^{\dagger}\left(0\right)\frac{\vec{\sigma}}{2}\psi\left(0\right)\right]^{2},\label{eq:localFLinteraction}\end{equation}
 where the prefactor is fixed such that the impurity susceptibility
is $\chi_{imp}=1/(4T_{K})$. This interaction lowers the ground state
energy when there is an odd number of remaining conduction electrons
($S_{el}=1/2$), thus splitting the charge steps in the strong coupling
limit. To lowest order in $1/T_{K}$, we find\begin{equation}
\frac{\mu_{\ell+\frac{1}{2}}-\mu_0^*}{\Delta}=\frac{[\ell-1]+1}{2}+(-1)^\ell\frac{\pi}{8}\frac{\xi_{K}}{L},\end{equation}
from which we obtain \begin{equation}
R\left(\frac{\xi_{K}}{L}\ll1\right)\approx\frac{\pi}{4}\frac{\xi_{K}}{L}.
\label{SC}
\end{equation}

\emph{Bethe Ansatz results}. -- 
\begin{figure}
\includegraphics[width=\columnwidth]{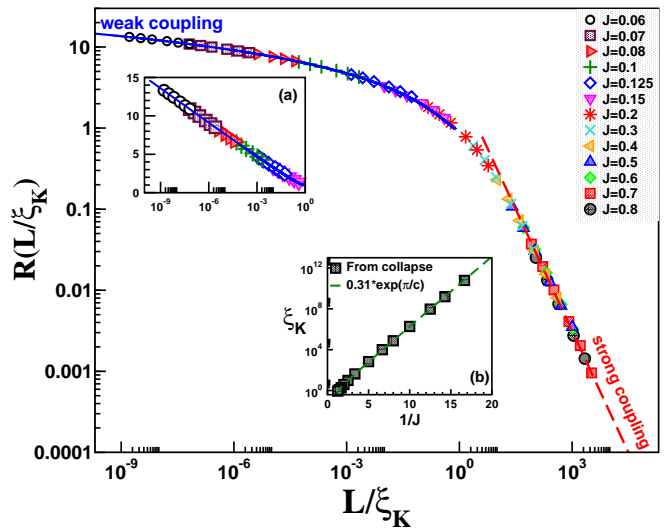}
\caption{(Color online) Universal ratio $R=\delta\mu^{\rm{odd}}/\delta\mu^{\rm{even}}$ as a scaling
function of $\xi_{K}/L$. Bethe ansatz results obtained for various systems sizes ($N_0=51,~101,~201,~501,~1001,~2001$) and 13 different values of the Kondo exchange $J$, indicated by different symbols. For each value of $J$, the system lengths $L$ have been rescaled $L\to L/\xi_K(J)$ in order to obtain the best collapse of the data using the strong coupling curve Eq.~(\ref{SC}) (dashed red line) as a support for the rest of the collapse. 
The weak coupling regime for $L\ll\xi_K$, enlarged in the inset (a), is described by the weak coupling expansion Eq.~(\ref{WC}) with ${\rm{constant}}\simeq 0.33$ (continuous blue curve).
Inset (b): The Kondo length scale, extracted from the universal data collapse of the main panel (black squares), is described by the exponential fit Eq.~(\ref{xi}) with $\xi_0\simeq 0.31$ (dashed green line).}
\label{BA}
\end{figure}
In order to check the scaling property of $R=\delta\mu^{\rm{odd}}/\delta\mu^{\rm{even}}$ for any $L/\xi_K$ we use the Bethe Ansatz (BA) solution of the one-channel Kondo problem~\cite{Andrei80}. We start with a half-filled
band of $N_0-1$ conduction electrons ($N_0$ odd) coupled to a localized impurity spin, corresponding 
to a system size $L=(N_0-1)/2$ in units where the Bethe ansatz cut-off parameter $D$ \cite{Andrei80}, 
related to the band-width, is set equal to one. 
Since the BA solution can be obtained for any filling factor, we can add particles one by one in the system ($N=N_{0},~N_{0}+1,~N_{0}+2,\hdots$) and compute the corresponding energies. This has been done  by solving numerically the coupled BA equations~\cite{Andrei80}, using a standard Newton-Rapson method.
In Fig.~\ref{BA} we present results obtained for the ratio 
$R=\delta\mu^{\rm{odd}}/\delta\mu^{\rm{even}}$ with $N_0=51,~101,~201,~501,~1001,~2001$ and 13 different values of the Kondo coupling in the range $0.06\le J\le 0.8$.
The universal scaling curve has been obtained by rescaling the x-axis $L\to L/\xi_K(J)$ in order to get the best collapse of the data. The entire crossover curve is obtained, ranging from the strong coupling $L\gg\xi_K$  to the weak coupling regime $L\ll\xi_K$. Both strong coupling [Eq.~\ref{SC})] and weak coupling [Eq.~(\ref{WC})] results are perfectly reproduced. As displayed in the inset (a) of Fig.~\ref{BA}, BA results agree with Eq.~(\ref{WC}) using only one fitting parameter: constant$\simeq 0.33$. The  Kondo length scale, shown in the inset (b) of Fig.~\ref{BA}, displays the expected exponential behavior~\cite{Andrei80}
\be
\xi_K(J)=\xi_0{\rm{e}}^{\pi/c},
\label{xi}
\ee
with $c=2J/(1-3/4J^2)$. Fitting $\xi_K$ to this expression gives $\xi_0\simeq 0.31$ (see inset (b) of Fig.~\ref{BA}) which is in very good agreement with the expected value of $\xi_0=1/\pi$, resulting~\cite{hewson} from the impurity susceptibility normalized to $1/(4T_K)$.

\begin{figure}
\includegraphics[%
  scale=0.34]{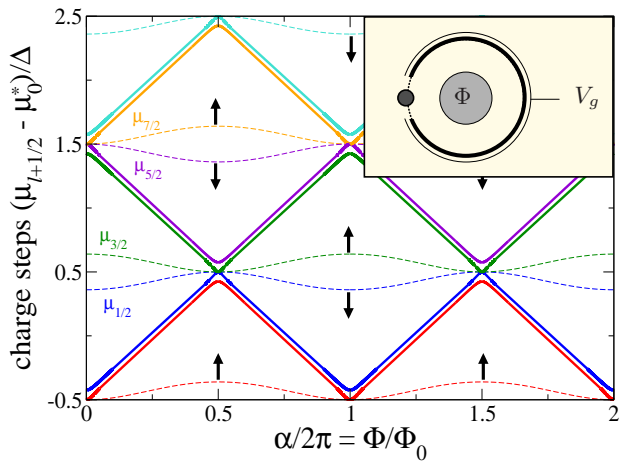}

\caption{(Color online) Flux dependence of the charge steps in the ring with
an embedded quantum dot (inset), assuming zero potential scattering. Dashed lines: weak coupling
regime ($\xi_{K}\gg L$); solid lines: strong coupling regime
($\xi_{K}\ll L$).\label{cap:(Color-online)-Flux}}
\end{figure}

\emph{Flux dependence in quantum rings.} -- The effect we have described
is not unique to the wire geometry. One interesting alternative is
to suppose that the quantum dot is embedded in a ring with
circumference $L$ (inset of Fig. \ref{cap:(Color-online)-Flux}).
This geometry offers the possibility of looking at the dependence
of the charge steps on the magnetic flux $\Phi$ threading the ring, which is related to the persistent current for the embedded quantum dot \cite{persistentcurrent,Zvyagin}.
 For simplicity, we neglect potential scattering and assume that the coupling between the dot and the leads is symmetric. In this case it is convenient to label the eigenstate
of the open chain by their symmetry under the parity transformation
$x\rightarrow L-x$. We define the even and odd fields $\psi_{e,o}\left(x\right)=[\psi_{R}\left(x\right)\pm e^{-ik_FL}\psi_{L}\left(L-x\right)]/\sqrt{2}$,
in terms of which the Kondo interaction takes the form \cite{persistentcurrent} \begin{eqnarray}
H_{K} & = & 4\pi v_{F}\lambda\left[\cos\frac{\alpha}{2}\psi_{e}^{\dagger}\left(0\right)-i\sin\frac{\alpha}{2}\psi_{o}^{\dagger}\left(0\right)\right]\nonumber
 \\
 &  & \times\frac{\vec{\sigma}}{2}\left[\cos\frac{\alpha}{2}\psi_{e}^{\phantom{\dagger}}\left(0\right)+i\sin\frac{\alpha}{2}\psi_{o}^{\phantom{\dagger}}\left(0\right)\right]\cdot\vec{S},\label{eq:Kondowithflux}\end{eqnarray}
where $\alpha\equiv2\pi\Phi/\Phi_{0}$ and $\Phi_{0}=hc/e$ is the
flux quantum. For $N_0=4p+1$ ($p$ integer), the lowest unoccupied state of the open chain ($\lambda=0$) is 
\emph{even} under parity. Defining $\delta \mu_n\equiv\mu_{2n+\frac{3}{2}}-\mu_{2n+\frac{1}{2}}$ as the splitting of the charge steps in the weak coupling limit, we find \begin{equation}
\frac{\delta\mu_n}{2\Delta}=\frac{3}{4}\left[1+\left(-1\right)^{n}\cos\alpha\right]\lambda_{eff}\left(L\right)+O\left(\lambda_{eff}^{2}\right).\end{equation}
Notice that for $\alpha=2m\pi$ only the even channel couples to the impurity and
the steps in the odd channel (corresponding to odd values of $n$)
do not split. The opposite happens when $\alpha=(2m+1)\pi$ (see Fig.
\ref{cap:(Color-online)-Flux}).%

In the strong coupling limit, the $\pi/2$ phase shift modifies the
boundary conditions in such a way that the transmission across the
dot becomes perfect \cite{delft,persistentcurrent}. The system is
then equivalent to an ideal 1D ring, whose eigenstates can be labeled
as right or left movers. The energy levels are flux dependent and
exhibit zigzag lines with level crossings at $\alpha=m\pi$ \cite{fuhrer}.  
The major effect of a Fermi liquid interaction analogous to Eq. (\ref{eq:localFLinteraction})
is to lift the degeneracy between right and left movers at $\alpha=m\pi$ and to split two out of the four values
of $\mu_{\ell+\frac{1}{2}}$ \cite{comment2}. We calculate the splitting $\delta\mu^\prime_n\equiv\mu_{2n+\frac{1}{2}}-\mu_{2n-\frac{5}{2}}$ using
degenerate perturbation theory to O($1/T_{K}$) and find ($\xi_{K}\ll L$) \begin{equation}
\frac{\delta\mu_{n}^\prime}{2\Delta}\approx\frac{\pi}{8}\frac{\xi_{K}}{L}+\sqrt{\left(\frac{\alpha-m\pi}{\pi}\right)^{2}+\left(\frac{\pi}{8}\frac{\xi_{K}}{L}\right)^{2}},\end{equation}
for $\alpha\approx m\pi, m=n,n\pm2,\dots$ (Fig. \ref{cap:(Color-online)-Flux}). This striking difference in
the flux dependence of the charge steps is consistent with the results in \cite{persistentcurrent}. 

Finally, we would like to point out that the constant interaction model assumed here neglects the exchange interaction between electrons in the wire. However, such interaction could be cast into one empirical parameter that gives a constant contribution to the even-odd splitting, even in the weak and strong coupling limits \cite{glazman}. 

In conclusion, we have shown that the width of the charge steps in a mesoscopic device depends on the finite size of the Kondo cloud. The signature of this effect would be the broadening of the Coulomb blockade valleys with total $N$ even as we increase $T_K$ (decrease $\xi_K$). The crossover between weak ($\xi_{K}\gg L$) and strong ($\xi_{K}\ll L$) coupling could be used to extract the characteristic length scale of the Kondo problem.

We thank E. S\o rensen and J. A. Folk for useful discussions. We acknowledge support 
from NSERC (RGP, NL, IA), CNPq (RGP, 200612/2004-2), CIAR (IA) and NSF (BIH, grant DMR-05-41988).


\begin{thebibliography}{10}
\bibitem{hewson}A. Hewson, \emph{The Kondo Effect to Heavy Fermions} (Cambridge University
Press, Cambridge, England, 1993).
\bibitem{glazman}For a review, see L.I. Glazman and M. Pustilnik, in: Nanophysics: Coherence and Transport.
Les Houches Session LXXXI, edited by H. Bouchiat \emph{et al}. (Elsevier,
Amsterdam, 2005). cond-mat/0501007
\bibitem{delft}W.G. van der Wiel \emph{et al}., Science \textbf{289}, 2105 (2000).
\bibitem{kondobox}W.B. Thim, J. Kroha, and J. von Delft, Phys. Rev. Lett. \textbf{82},
2143 (1999);  H. Hu, G.-M. Zhang, and L. Yu, Phys. Rev. Lett. \textbf{86}, 5558 (2001); P.S. Cornaglia and C.A. Balseiro, Phys. Rev. B \textbf{66}, 115303 (2002).
\bibitem{persistentcurrent}I. Affleck and P. Simon, Phys. Rev. Lett. \textbf{86}, 2854 (2001);
P. Simon and I. Affleck, Phys. Rev. B \textbf{64}, 085308 (2001).
\bibitem{key-4}P. Simon, J. Salomez, and D. Feinberg, Phys. Rev. B \textbf{73}, 205325
(2006); R.K. Kaul \emph{et al}., Phys. Rev. Lett. \textbf{96}, 176802
(2006).
\bibitem{goldhaber-gordon}R.M. Potok \emph{et al}., cond-mat/0610721.
\bibitem{comment}We can account for particle-hole symmetry breaking by adding to the
Hamiltonian a marginal scattering potential term with amplitude $\sim O\left(J\right)$.
This term simply shifts the position of the charge steps, without
affecting $R$. 
\bibitem{nozieres}P. Nozi\`eres, J. Low. Temp. Phys. \textbf{17}, 31 (1974).
\bibitem{Andrei80}  N. Andrei, Phys. Rev. Lett. \textbf{45}, 379 (1980); P.B.
Wiegman, JETP Lett., \textbf{31}, 364 (1980); N. Andrei, K. Furuya and J.H. Lowenstein, Rev. Mod. Phys. {\bf 55}, 331 (1983). 
\bibitem{Zvyagin}A.A. Zvyagin and T.V. Bandos, Low Temp. Phys. \textbf{20}, 222 (1994); K. Kang and S.-C. Shin, Phys. Rev. Lett. {\bf85}, 5619 (2000); H.-P. Eckle, H. Johannesson, and C.A. Stafford, Phys. Rev. Lett. \textbf{87}, 016602 (2001); A.A. Aligia, Phys. Rev. B {\bf66}, 165303 (2002).
\bibitem{fuhrer}A. Fuhrer \emph{et al}., Nature (London) \textbf{413}, 822 (2001).
\bibitem{comment2} In the more general case of nonzero but weak potential scattering one has avoided level crossings at $\alpha=m\pi$ and the double steps for small $\xi_K/L$ in Fig. \ref{cap:(Color-online)-Flux} also split.
\end{thebibliography}
\end{document}